\documentclass[aps,prl,preprint,superscriptaddress]{revtex4-1}

\usepackage{graphicx}
\usepackage[final]{pdfpages}

\begin{document}


\title{Dynamics and dissipation induced by single-electron tunneling in carbon nanotube nanoelectromechanical systems}


\author{Marc Ganzhorn}
\author{Wolfgang Wernsdorfer} 
\affiliation{Institut N\'eel, CNRS \& Universit\'e Joseph Fourier, BP 166, 25 Avenue des Martyrs, 38042 Grenoble Cedex 9, France}


\date{\today}

\begin{abstract}
We demonstrate the effect of single-electron tunneling (SET) through a carbon nanotube quantum dot on its nanomechanical motion. 
We find that the frequency response and the dissipation of the nanoelectromechanical system (NEMS) to SET strongly depends on the 
electronic environment of the quantum dot, in particular on the total dot capacitance and the tunnel coupling to the metal contacts. 
Our findings suggest that one could achieve quality factors of 10$^{6}$ or higher by choosing appropriate gate dielectrics 
and/or by improving the tunnel coupling to the leads. 
\end{abstract}



\maketitle


Carbon nanotubes (CNT) have become an essential building block for nanoelectromechanical systems (NEMS). Their low mass and high Young's modulus allow for instance ultrasensitive mass~\cite{Lassagne2008,Chiu2008,Jensen2008} or force detection~\cite{Eichler2011,Lassagne2011} (electric and magnetic) over a wide range of frequencies and its small diameter enables even single-molecule detection~\cite{Lassagne2011,Bogani2008,Urdampilleta2011}. Moreover, CNT devices exhibit remarkable electronic transport properties, ranging from Kondo physics~\cite{Nygard2000} to Coulomb blockade at high temperature~\cite{Charlier2007}. It was recently demonstrated that a CNT NEMS' nanomechanical motion at very low temperature (i.e. in Coulomb blockade regime) is strongly affected by the electronic transport through the CNT quantum dot (QD), and vice versa: For instance, single-electron tunneling (SET) caused a frequency softening and increased dissipation when tuning the CNT dot's potential through a Coulomb peak~\cite{Lassagne2009,Steele2009,Huttel2009}.  

Here we demonstrate that the response and dissipation of a CNT NEMS at very low temperature induced by zero bias SET through the CNT NEMS-QD critically depends on the dot capacitance, the tunnel coupling to the metal leads and temperature. We studied the frequency and dissipation response of nanomechanical bending modes to zero bias SET in suspended CNT devices with tunable tunnel couplings and different gate dielectrics, i.e. different dot capacitance. We observe that SET causes a frequency softening for small dot capacitance and/or tunnel coupling, whereas a frequency hardening or no frequency modulation is observed for large dot capacitance and/or tunnel coupling. We show that the dissipation of the CNT NEMS is mainly dominated by the capacitance, when electron tunneling through the dot is suppressed (i.e. in the Coulomb valley), whereas it is limited by the mean tunneling rate $\Gamma$ and the conductance, when electron tunneling through the dot is allowed (i.e. on a Coulomb peak). Finally we demonstrate that the tunnel current is the dominant dissipation mechanism in CNT NEMS at low temperature. Our findings are in fair agreement with a theoretical model provided previously~\cite{Lassagne2009,Steele2009}.

\begin{figure}
\includegraphics[width = 14cm]{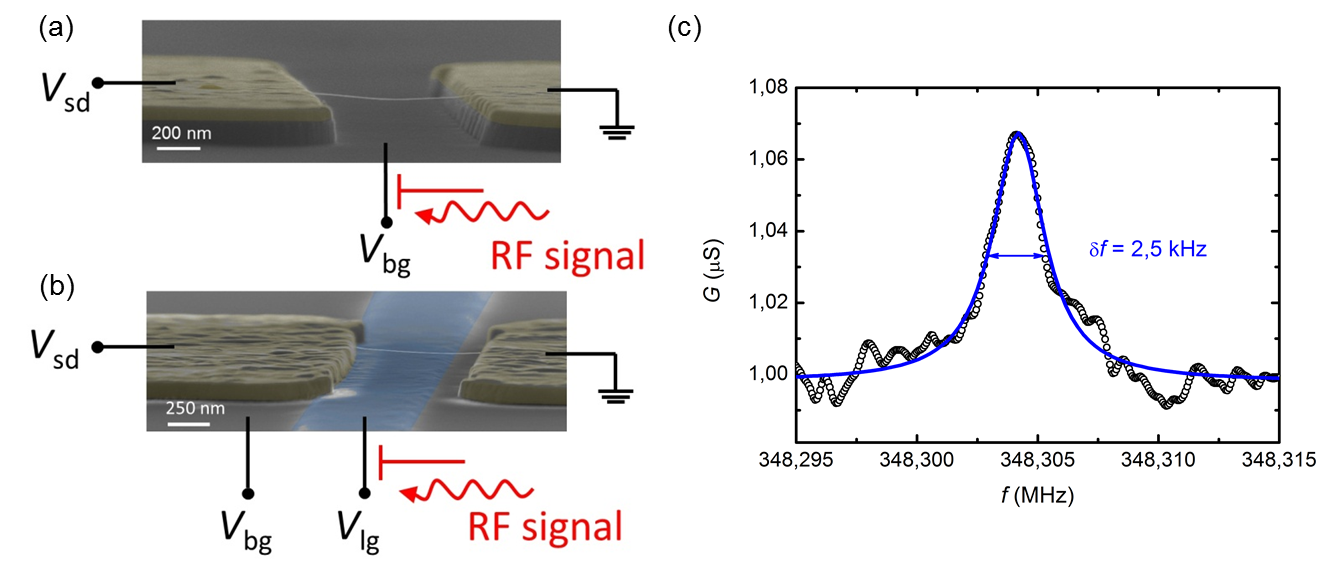}
\caption{False color SEM image of (a) a low capacitance device based on a SiO$_2$ covered Si backgate (grey color) and (b) a high capacitance device based on an Al$_2$O$_3$ covered local metallic gate (blue). The RF actuation signal is injected into the local metallic gate and Si backgate, respectively, through a home-built bias-T. As the induced mechanical motion changes the charge flow through the CNT quantum dot and vice versa, we can detect the CNT resonance through a change in zero bias conductance. (c) Mechanical resonance of a typical CNT NEMS at low driving power $P_{\text{RF}} = -100$ dBm. The resonance width $\delta f = 2.5$ kHz leads to a quality factor of $Q \approx 140000$. }
\label{fig:fig1}
\end{figure}

Low capacitance CNT quantum dots ($C_{\text{dot}}$ $\approx$ 20 - 40 aF) were obtained by using silicon dioxide as gate dielectric [Fig. 1(a)]. First, source-drain electrodes are patterned by optical DUV lithography and e-beam evaporation of Mo (20 nm) and Pt (160 nm) on 500 nm of thermal SiO$_2$. To ensure the suspension of the CNT, 150 nm of SiO$_2$ are dry etched in CHF$_3$ plasma. High capacitance CNT quantum dots ($C_{\text{dot}}$ $\approx$ 160 - 260 aF) are obtained with high-$\kappa$ gate dielectrics Al$_2$O$_3$ [Fig. 1(b)]. First, a 1 $\mu$m-wide metallic local gate is patterned by optical DUV lithography and e-beam evaporation of Mo (20 nm) on 300 nm of thermal SiO$_2$. A layer of 100 nm of Al$_2$O$_3$ is then deposited by atomic layer deposition. Using optical DUV lithography and e-beam evaporation of Mo (20 nm) and Pt (160 nm), source-drain electrodes are aligned above the local gate. Suspended CNT's are finally grown by chemical vapor deposition at 800$^{\circ}$C from a CH$_4$ feedstock and Fe/Mo catalyst spots patterned on the source-drain electrodes next to the junction. The CNT device length is approximately 800 nm ($\pm$ 50 nm) and the dot capacitances are deduced from the dot's charging energy at low temperature. The spread in capacitance values of nominally identically fabricated devices is due to variations in the nanotubes length and slack (see Supporting Information).

The measurements are carried out in a $^3$He/$^4$He dilution refrigerator with a base temperature of 30 mK. The NEMS actuation and detection scheme used in our experiment is similar to the one used by Steele and co-workers~\cite{Steele2009,Huttel2009}. The RF actuation signal is injected into the gate electrode via a home-built bias T. As the induced mechanical motion changes the charge flow through the CNT quantum dot and vice versa, we can detect the CNT resonance through a change in zero bias conductance. The actuation power is kept to a minimum ($P_{\text{RF}}$ $\approx$ -100 dBm) in order to ensure a quasi-linear regime of the oscillator and a high $Q$ lorentzian resonance shape [Fig. 1(c)]. All measurements were done under zero bias with a standard lock-in technique. 

SET in a CNT NEMS-QD can be considered as an external perturbation to the CNT mechanical motion, and vice versa. This perturbation can be described as an electrodynamic force acting on the CNT. The contribution of this force, which is in phase with the mechanical motion, is responsible for a frequency modulation $\Delta f$, whereas a contribution of the force being out of phase with mechanical motion induces a modification of the dissipation and the quality factor $Q$. $\Delta f$ and $Q$ can be expressed as follows~\cite{Lassagne2009}

\begin{eqnarray}
	\Delta f &=& - \frac{f_{\text{0}}}{2}\frac{C_{\text{g}}^{'2}}{k}\frac{V_{\text{g}}^{2}}{C_{\text{dot}}} \left( \frac{2G}{C_{\text{dot}}\Gamma} -1 \right) \\
	\frac{1}{Q} &=& 2\pi f\frac{C_{\text{g}}^{'2}}{k} V_{\text{g}}^{2} \left(\frac{2}{{\Gamma} C_{\text{dot}}} \right)^2 G
\end{eqnarray}

Where $\Gamma$ is the mean tunnel coupling of the dot to the leads, yielding $\Gamma = \left( \Gamma_{\text{d}}+\Gamma_{\text{s}} \right)/2$, and $C_{\text{dot}}$ is the total dot capacitance. We can estimate the tunnel coupling $\Gamma$ for each device from the line shape of the Coulomb blockade peaks (see Supporting Information). Eqs. (1) and (2) are valid in the regime of Coulomb blockade and if $\Gamma \gg f_0$. 

\begin{figure*}
\includegraphics[width = 16cm]{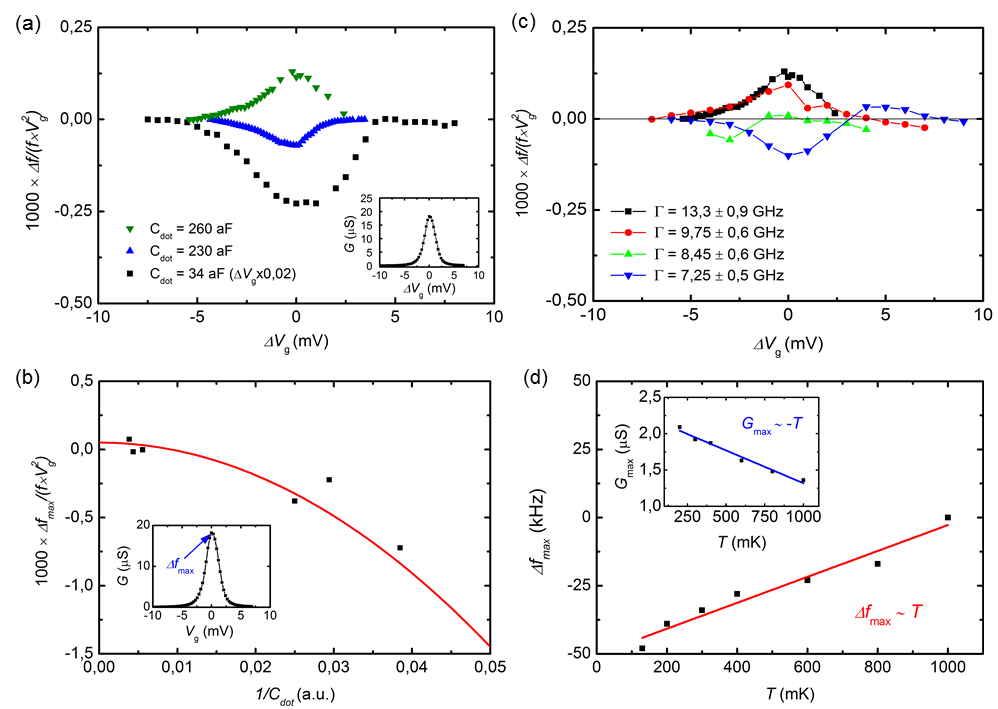}
\caption{Modulation of the resonance frequency due to SET: 
(a) Frequency shift $\Delta f$ (scaled with $f \cdot V_{g}^2$) for different dot capacitance $C_{\text{dot}}$, with comparable tunneling rates $\Gamma$. inset shows a typical Coulomb peak centered at $\Delta V_{\text{g}} = 0$; (b) Maximum scaled frequency shift at maximum tunnel current $G_{\text{max}}\left(\Delta V_g = 0 \right)$ vs. the dot capacitance for all measured devices. The solid red line corresponds to the fit using eq. 1 with $C_{\text{g}}^{\text{'2}}/k$ = 8 $\cdot$ 10$^{\text{-22}}$ F$^2$/Nm and $\Gamma_{\text{d}} \approx$ 10 GHz. The data are in fair agreement with the model. (c) Scaled frequency shift for different Coulomb peaks i.e. tunneling rates $\Gamma$, with a fixed dot capacitance $C_{\text{dot}} = 260$ aF (device 1). The tunneling rate $\Gamma$ is estimated from the lineshape of the respective Coulomb peaks (see Supporting Information). (d) Maximum frequency shift $\Delta f_{\text{max}}$ at maximum tunnel current $G_{\text{max}}\left(\Delta V_g = 0 \right)$ vs. temperature for a device yielding $C_{\text{dot}} = 160$ aF (device 2). The inset shows the evolution of $G_{\text{max}}$ as a function of temperature.}
\label{fig:fig2}
\end{figure*}

First, we compare the frequency response $\Delta f$ (scaled with $f \cdot V_{g}^2$) of devices with different dot capacitances $C_{\text{dot}}$. We studied the frequency response for comparable Coulomb peaks height, i.e. for comparable conductance $G\left(V_{\text{g}}\right) \approx$ 20 $\mu$S. The tunnel coupling are on the order of 10 GHz. As depicted in Fig. 2(a), we observe a strong frequency softening for small dot capacitances, whereas for high dot capacitances a frequency hardening or no modulation is observed. As shown in Fig. 2(b), the experimental data are in fair agreement with the model. A fit with Eq. 1 yields fitting parameters $C_{\text{g}}^{\text{'2}}/k$ = 8 $\cdot$ 10$^{\text{-22}}$ F$^2$/Nm and $\Gamma \approx$ 10 GHz, which are in rather good agreement with previous experiments~\cite{Lassagne2009,Steele2009,Sazonova2004}. Fig. 2(c) shows the (scaled) frequency modulation for different Coulomb peaks and tunneling rates  $\Gamma$ on a given device (device 1). It was shown previously that one can tune a CNT quantum dot from a Coulomb blockade regime to a regime of strong tunnel coupling by simply changing the CNT's gate potential~\cite{Jespersen2011,Steele2009a}. Hence, we can change the tunnel coupling $\Gamma$ by tuning the gate voltage of our CNT device and estimate the values of $\Gamma$ from the Coulomb peak line shape (see Supporting Information). The dot capacitance $C_{\text{dot}}$ = 260 aF remain unchanged by tuning the gate voltage. As $\Gamma$ increases, the CNT becomes stiffer, resulting in a sign change of $\Delta f$ and the transition from a frequency softening to frequency hardening, in qualitative agreement with Eq. 1 [Fig. 2(c)].

\begin{figure*}
\includegraphics[width = 16cm]{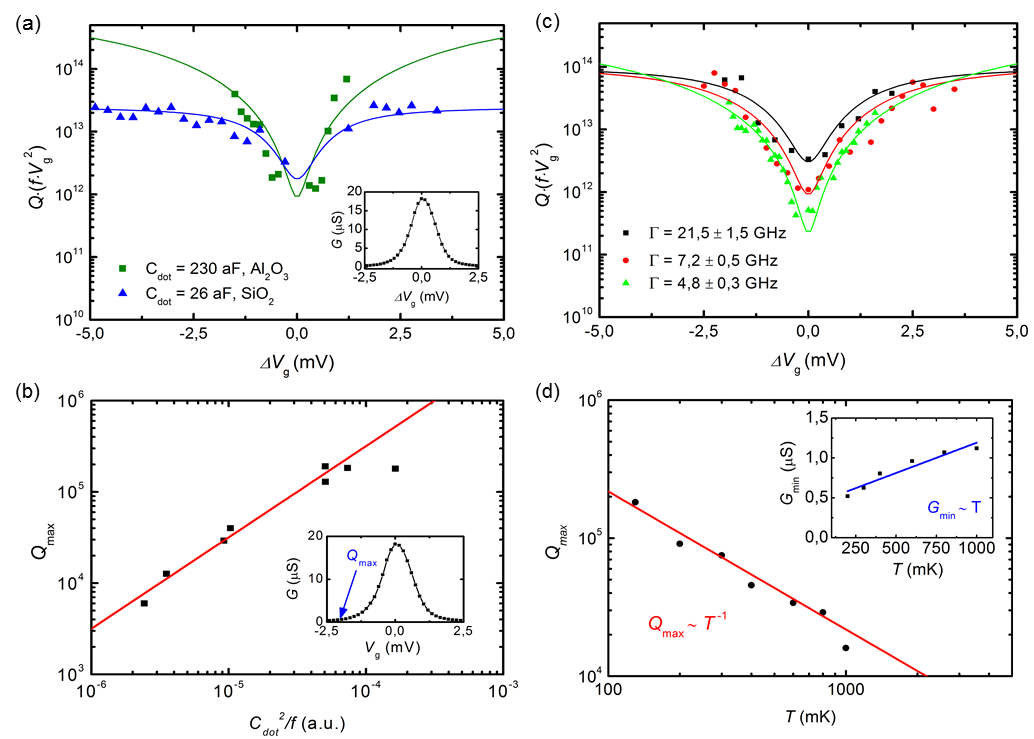}
\caption{Modulation of the $Q$-factor (scaled with $f \cdot V_{g}^2$) due to SET: (a) For different dot capacitances, with comparable tunneling rates $\Gamma$ . The solid lines are guide to the eye and the inset shows a representative Coulomb peak centered at $\Delta V_{\text{g}} = 0$. (b) Maximum quality factor $Q_{\text{max}}$ at minimal tunnel current $G\left(|\Delta V_{\text{g}}| \gg 0\right)$ vs. the dot's capacitance for all measured devices. The solid red line represents the fit with eq. 2 with $C_{\text{g}}^{\text{'2}}/k$ = 1 $\cdot$ 10$^{\text{-21}}$ F$^2$/Nm and $\Gamma \approx$ 10 GHz. Our findings are in good agreement with the model. (c) Modulation of the $Q$-factor (scaled with $f \cdot V_{g}^2$) for different tunneling rates $\Gamma$, with a fixed dot capacitance $C_{\text{dot}} = 160$ aF (device 2). The solid lines are guide to the eye and the tunnel coupling $\Gamma$ is estimated from the lineshape of the respective Coulomb peaks (see Supporting Information). (d) Maximum quality factor $Q_{\text{max}}$ at minimal tunnel current $G\left(|\Delta V_{\text{g}}| \gg 0\right)$ vs. temperature for a device yielding $C_{\text{dot}} = 160$ aF (device 2). The inset shows the evolution of $G_{\text{min}}$ as a function of temperature. }
\label{fig:fig3}
\end{figure*}

Fig. 3(a) depicts the effect of SET on the dissipation of the CNT NEMS, i.e. the $Q$ factor (scaled with $f \cdot V_{g}^2$), for devices with different dot capacitances. From Eq. 2, we expect an increase of the Q factor with increasing capacitance in the limit of suppressed SET through the dot. Indeed, in a region of suppressed SET ($|\Delta V_{\text{g}}| \gg 0$),the $Q$ factor is larger for devices with higher dot capacitance [Fig. 3(a)]. The tunnel coupling $\Gamma \approx 10$ GHz and the conductance $G\left(V_{\text{g}}\right) \approx$ 18 $\mu$S are comparable for both traces. Fig. 3(b) shows the $Q$ factor (scaled with $f \cdot V_{g}^2$) in the limit of suppressed SET ($|\Delta V_{\text{g}}| \gg 0$) as a function of the dot capacitance for all measured devices. The fit with Eq. 2 yields $C_{\text{g}}^{\text{'2}}/k$ = 1 $\cdot$ 10$^{\text{-21}}$ F$^2$/Nm and $\Gamma \approx$ 10 GHz which is consistent with the fitting parameters of Fig. 2(b) and previous experiments~\cite{Lassagne2009,Steele2009,Sazonova2004}. Despite the good agreement between theory and experiment, we observe deviations from the model for some devices in Fig. 2(b) and 3(b), which we attribute to variations in $C_{\text{g}}^{\text{'2}}/k$ (see Supporting Information). Moreover, we expect from Eq. 2 and previous experiments~\cite{Lassagne2009,Steele2009} an enhanced electromechanical dissipation in a given device if the conductance $G$ (i.e. the tunnel current) and the tunnel resistance at the nanotube-electrode interface $R \sim 1/ \Gamma_{\text{d}}$ are increased, in analogy to the current dissipation in a simple resistance. Indeed we observe increased dissipation (decreased Q factor) when tuning the gate voltage through a Coulomb peak [Fig. 3(a) and 3(c)]. In Fig 3(c) (device 2), this effect becomes more pronouced as we move to gate voltage regions with a smaller tunneling rate $\Gamma$, i.e. larger tunnel resistance at the nanotube-electrode interface. The dot capacitance $C_{\text{dot}}$ yields 160 aF for this device and the height of the Coulomb peak are comparable for all traces ($G \approx$ 7 $\mu$S). 
Therefore, the CNT capacitance is the limiting factor for the dissipation in regions of suppressed SET, whereas the mean tunnel coupling and the tunneling current itself define the dissipation in regions of strong SET. We conclude that the electron transport through the CNT is the main dissipation mechanism in our nanoelectromechanical system. 

Finally we study the temperature dependance of the frequency response and the dissipation. Fig. 2(d) depicts the frequency shift $\Delta f_{\text{max}}$ for $G_{\text{max}} = G\left(\Delta V_{\text{g}} = 0\right)$ whereas Fig. 3(d) shows the quality factor $Q_{\text{max}}$ for $G_{\text{min}} = G\left(|\Delta V_{\text{g}}| \gg 0\right)$ as function of temperature. A fit of the data yields $\Delta f_{\text{max}} \sim T$ and $Q_{\text{max}} \sim 1/T$. As temperature increases the Coulomb blockade peak broadens and becomes smaller (see supporting information  Fig. S3). As a result $G_{\text{max}} = G\left(\Delta V_{\text{g}} = 0\right)$ decreases whereas $G_{\text{min}} = G\left(|\Delta V_{\text{g}}| \gg 0\right)$ increases with temperature, as depicted in the insets of Figs. 2(d) and 3(d).  From the data fit we obtain $G_{\text{max}} \sim -T$ and  $G_{\text{min}} \sim T$. It was previously demonstrated that carbon nanotubes show Luttinger liquid behavior~\cite{Bockrath1999,Kane1997,Egger1997}, where the conductance follows a characteristic power law dependence as a function of temperature $G(T) \sim T^{\alpha}$, $\alpha$ being related to the Luttinger parameter g by $\alpha =(g-1+g-2)/2$~\cite{Bockrath1999,Kane1997}. We find $\alpha \approx 1$ and a Luttinger parameter of $g = 0,1$, which is close to previous measurements~\cite{Bockrath1999} and theoretical predictions~\cite{Kane1997,Egger1997}. Comparing the temperature dependance of the conductance, the quality factor and the frequency shift we finally obtain $\Delta f_{\text{max}} \sim -G_{\text{max}}$ and $Q_{\text{max}} \sim 1/G_{\text{min}}$, which is in full agreement with Eqs. 1 and 2. We can conclude that the current is the dominant dissipation mechanism for carbon nanotube based NEMS in a Coulomb blockade regime at cryogenic temperatures.

We have demonstrated that the response and the dissipation to single-electron tunneling of carbon nanotube NEMS at low temperature depends on the dot capacitance and the tunnel coupling to the leads, the tunnel current itself being the main dissipation mechanism in the system. By choosing high-$\kappa$ dielectrics (HfO$_2$ or ZrO$_2$) and improving the tunnel contact to the metal leads, one could significantly enhance the quality factor $Q$ to values exceeding 10$^6$. It was proposed in theoretical calculation that one can use such high$-Q$ CNT NEMS as magnetic torque or force detectors for nanoparticles~\cite{Lassagne2011} or single molecule magnets~\cite{Kovalev2011,Garanin2011} grafted to the CNT NEMS. In principle one can achieve a sensitivity of one $\mu_{\text{B}}$ at low temperature, whereas the best magnetometers, for instance the micro-SQUID, only have a sensitivity of 10$^3$ $\mu_{\text{B}}$~\cite{Wernsdorfer2009}. Finally, such molecular quantum spintronic device would allow single spin manipulation on a molecular level.

\begin{acknowledgments}
This work is partially supported by the ANR-PNANO project MolNanoSpin No. ANR-08-NANO-002 and ERC Advanced Grant MolNanoSpin No. 226558. M.G. acknowledges the financial support from the RTRA Nanosciences Foundation. Samples were fabricated in the NANOFAB facility of the Neel Institute. We thank F. Balestro, E. Bonet, T. Crozes, J.P. Cleuziou, E. Eyraud, T. Fournier, R. Haettel, C. Hoarau, D. Lepoittevin,  V. Nguyen, V. Reita, A. Reserbat-Plantey, C. Thirion, M. Urdampilleta.
\end{acknowledgments}



%

\appendix*
\newpage


\section*{Supplementary material}

# Dynamics and dissipation induced by single-electron tunneling in carbon nanotube nanoelectromechanical systems
# Supporting Information

M. Ganzhorn, W. Wernsdorfer

Institut Neel, CNRS & Universite Joseph Fourier, BP 166,
25 Avenue des Martyrs, 38042 Grenoble Cedex 9

**Content**

1. Estimating quantum dot capacitances of CNT NEMS
2. Estimating the tunnel coupling $\Gamma$ in CNT NEMS
3. Accuracy of fitting parameter $C_g'^2/k$
4. Power law temperature dependence of zero bias conductance

# 1. Estimating quantum dot capacitances of CNT NEMS

The total dot capacitance is obtained from bias spectroscopy measurements. The total dot capacitance $C_{dot}$ is related to the charging energy $E_c$ of the dot via $C_{dot} = e/E_c$. The charging energy is deduced from the size of the Coulomb diamonds (fig. S1a). Alternatively one can determine the capacitances $C_s$ and $C_d$ from the slope of the diamond edges and the gate capacitance $C_g$ from the distance between 2 degeneracy points (Fig S1b). The total dot capacitance then yields $C_{dot} = C_s + C_d + C_g$.

For $SiO_2$ based devices we obtain capacitance between 20 and 40 aF, whereas for $Al_2O_3$ based devices the values range from 160 aF to 260 aF. The spread in capacitance values for nominally identically fabricated devices can mainly be attributed to different device lengths (ranging from 700 nm to 850 nm) and therefore to variations in charging energies. Electron microscopy revealed slack on most devices, suggesting a low built-in tension which is characteristic for CVD grown CNT's.[1,2] As the slack may be a different for each device, the distance between the CNT and the gate can vary from one device to another, which would also contribute to the spread in capacitance values.

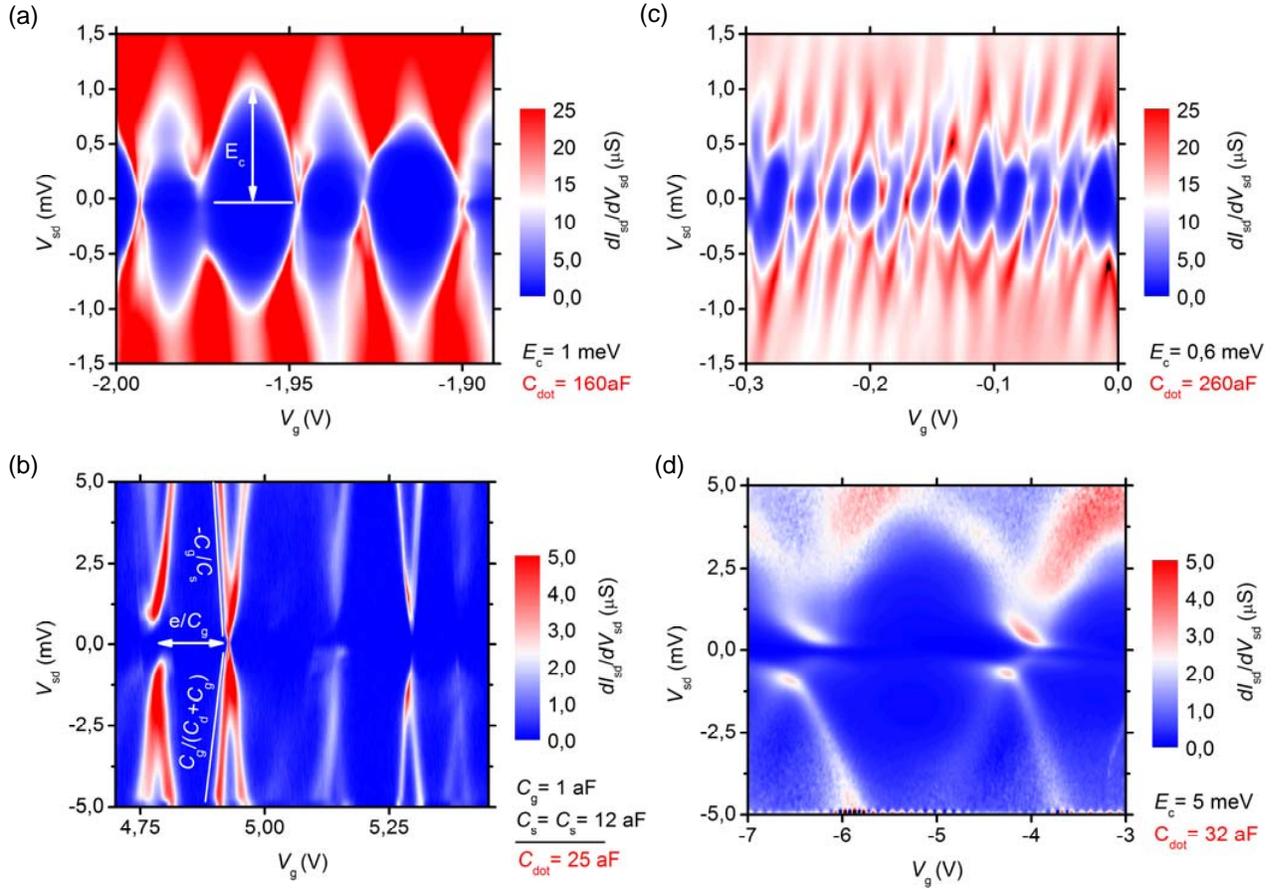

**Fig. S1: Estimating the dot capacitance via bias spectroscopy:** (a)(c) Coulomb diamonds for two $Al_2O_3$ based devices with a length of 780 nm and 820 nm, respectively. (b)(d) Coulomb diamonds for two $SiO_2$ based devices.

## 2. Estimating the tunnel coupling Γ in CNT quantum dot

In the regime of Coulomb blockade, the energy level broadening in the carbon nanotube quantum dot is given by the electronic temperature and the tunnel coupling to the leads. The electronic temperature in our system is around 150 mK, therefore at least an order of magnitude smaller than the expected tunnel couplings in carbon nanotube junctions Γ>> kT. In this limit, the Coulomb peak line shape can approximated as follows[3] :

$$\frac{G}{G_{max}} = \frac{(\Gamma/2)^2}{\alpha^2(V_g - V_{g,0})^2 + (\Gamma/2)^2} \quad (S1)$$

Where $G_{max}$ is the maximum of the Coulomb peak, $\Gamma = (\Gamma_s + \Gamma_d)/2$ the mean tunneling rate through the quantum dot, $V_{g,0}$ the center of the peak and α an energy scaling factor given by:

$$\alpha = \frac{1}{1.36}\frac{dV_{sd}}{dV_g} = \frac{1}{1.36}\frac{C_g}{C_s + C_d + C_g} \quad (S2)$$

We used Eq. S1 to fit the Coulomb peaks and estimate the mean tunnel coupling Γ for devices 1 and 2 studied in Fig. 2 (c) and Fig. 3 (c) of the main text. The results are depicted in Fig. S2 with the same color code than in the main text figures. For device 1, studied with respect to the frequency shift *Δf*, we can tune the tunnel coupling Γ from 30 μeV to 55 μeV (7.25 GHz to 13.3 GHz) [Fig. S2 (a)], leading to a stiffening of the CNT [Fig. 2(c)]. For device 2, studied with respect the *Q* factor modulation, we can change Γ from 20 μeV to 90 μeV (4.8 GHz to 21.5 GHz), leading to an increase of the quality factor [Fig. 3(c)].

Although Eq. (1) is only an approximation (with an error estimated to 7%), it is sufficiently accurate to show a dependence of the frequency shift and the dissipation on the tunnel coupling in <u>qualitative</u> agreement with Eq. 1 and 2 of the main text.

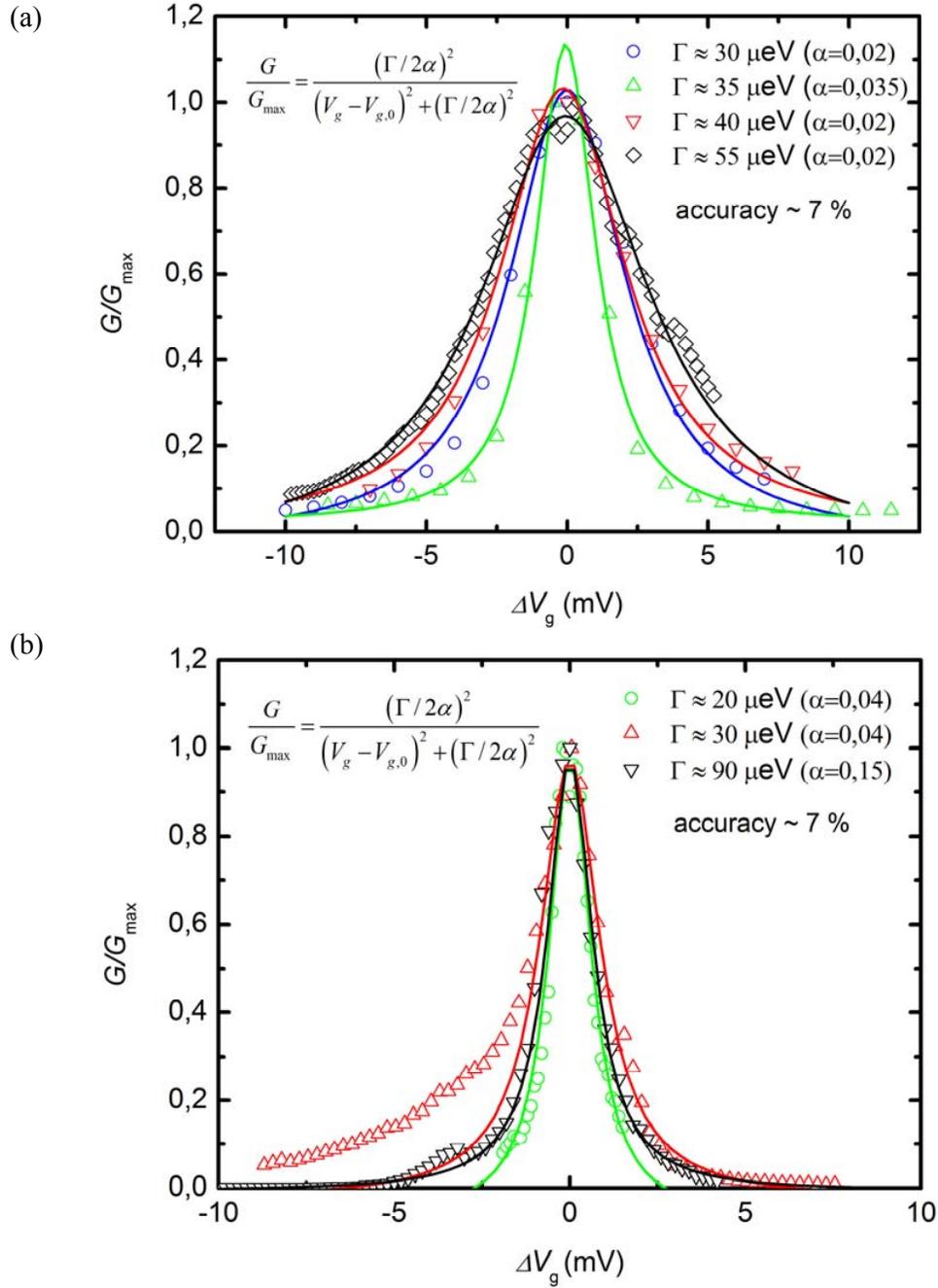

**Fig. S2: Estimating the tunnel coupling Γ in the limit of tunnel broadened Coulomb blockade.** a) Device 1 studied in Fig. 2(c) of the main text with respect to the frequency shift Δf. b) Device 2 studied in Fig. 3(c) of the main text with respect to the quality factor Q. The color code is identical to the one of the main text figures. The estimation is accurate enough to

4

demonstrate a qualitative agreement of our data in Fig. 2 (c) & 3 (c) with the model and Eq. 1 & 2 of the main text.

## 3. Accuracy of fitting parameter $C_g^{'2}/k$

In Fig. 2b and 3b of main text, we fitted the experimental data with equation 1 and 2, using $\Gamma$ and $C_g^{'2}/k$ as fitting parameters. We obtain average value for $C_g^{'2}/k = 10^{-21}$ F$^2$/Nm and $\Gamma = 10$ GHz, which is in fair agreement with previous experiments.[4,5] Although the model fits the experimental data with good accuracy, we can observe certain deviation from the model for some devices, which could indeed be attributed to variations in $C_g^{'2}/k$:

According to ref. 4, $C_g^{'2}/k$ is given by $C_g^{'2}/k = (C_g/Z)^2/k$ in good approximation, with $C_g$ being the gate capacitance, $Z$ the CNT-gate distance and $k$ the CNT's spring constant. We studied two different device layouts: SiO$_2$ based CNT QD with a gate capacitance $C_g = 2\pm1$ aF and $Z = 700$ nm and Al$_2$O$_3$ based CNT QD with $C_g = 3\pm1$ aF and $Z = 300$ nm. $C_g$ is estimated from the voltage separation of two adjacent Coulomb peaks (see Fig. S1) and $Z$ is given by the sum of the oxide thickness and thickness of the metal electrodes. With an average spring constant $k = 10^{-3}$ N/m (value extracted from ref. 4 and 5) we obtain values for $C_g^{'2}/k = (6\pm2)\cdot10^{-21}$ F$^2$/Nm, which is close to values obtained from the fit in Fig. 2b and 3b.

The above estimation shows that we indeed have variations in $C_g^{'2}/k$, caused by a small spread in capacitance values $C_g$ and CNT-gate distance $Z$, which could ultimately lead to deviations of some devices from the model fit.

## 4. Power law temperature dependence of the zero bias conductance

As temperature increases the Coulomb blockade peak broadens and becomes smaller (see Fig. S2). As a result $G_{max}= G(\Delta V_g = 0)$ decreases whereas $G_{min}= G(|\Delta V_g| = -1$ mV) increases with temperature, which is depicted in the insets of Figs. 2(d) and 3(d) of the main text. Both follow a power law dependence, for instance $G_{max} \sim -T$ and $G_{min} \sim T$, as expected for Luttinger liquids like carbon nanotubes. [6,7,8]

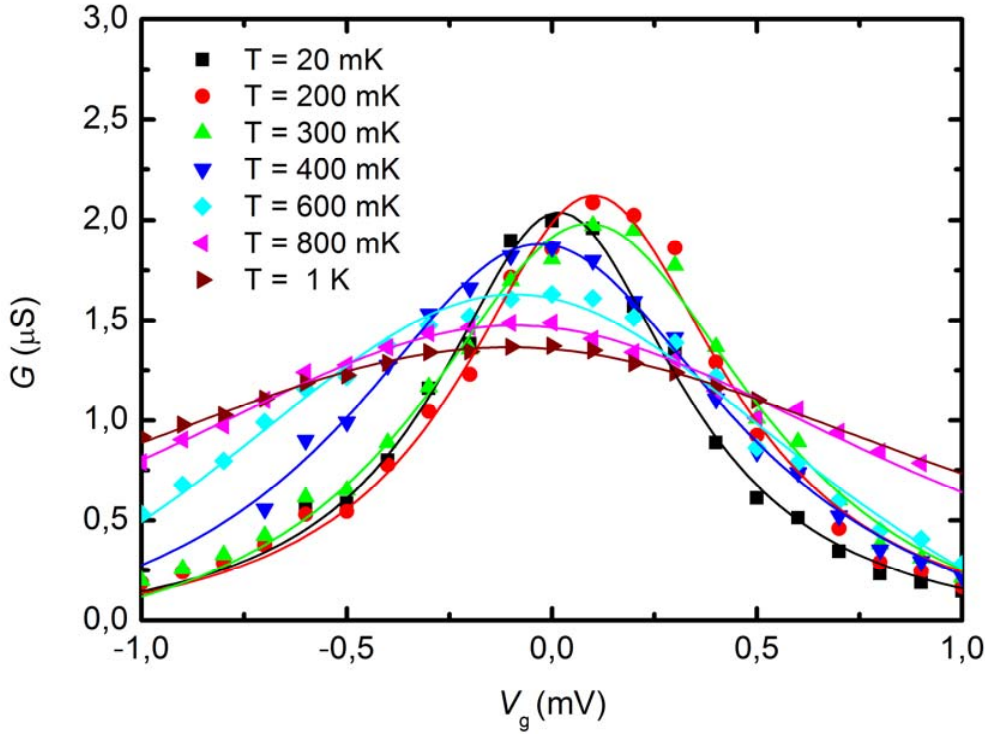

**Fig. S3: Evolution of the zero-bias conductance as a function of temperature** (from 20 mK to 1 K) for the device studied in Fig. 2(d) and 3(d) of the main text.

---



\end{document}